\documentclass[12pt]{article}
\usepackage{graphicx}

\newcommand\pubnumber{DPF2013-57}
\newcommand\pubdate{\today}

\def\napoli{Argonne National Laboratory\\
Iowa State University, USA}
\def\Title#1{\begin{center} {\Large #1 } \end{center}}
\def\Author#1{\begin{center}{ \sc #1} \end{center}}
\def\Address#1{\begin{center}{ \it #1} \end{center}}

\newcommand\pubblock{\rightline{\begin{tabular}{l} \pubnumber\\
         \pubdate  \end{tabular}}}
\newenvironment{Abstract}{\begin{quotation}  }{\end{quotation}}
\newenvironment{Presented}{\begin{quotation} \begin{center} 
             PRESENTED AT\end{center}\bigskip 
      \begin{center}\begin{large}}{\end{large}\end{center} \end{quotation}}
\def\Acknowledgments{\bigskip  \bigskip \begin{center} \begin{large}
             \bf ACKNOWLEDGMENTS \end{large}\end{center}}
\def\beq{\begin{equation}}
\def\eeq#1{\label{#1}\end{equation}}
\def\eeqn{\end{equation}}

\def\beqa{\begin{eqnarray}}
\def\eeqa#1{\label{#1}\end{eqnarray}}
\def\eeqan{\end{eqnarray}}

\let\bar=\overbar

\def\Dslash{\not{\hbox{\kern-4pt $D$}}}
\def\dslash{\not{\hbox{\kern-2pt $\del$}}}

\def\msb{{\bar{\ssstyle M \kern -1pt S}}}

\begin{document}
\begin{titlepage}
\pubblock

\vfill
\Title{Using Fast Photosensors in Water Cherenkov Neutrino Detectors}
\vfill
\Author{ Ioana Anghel}%\support}
\Address{\napoli}
\vfill
\begin{Abstract}
Many of the yet unanswered questions in neutrino physics, such as CP violation in the lepton sector or neutrino mass hierarchy, could be answered with higher sensitivity neutrino experiments. New photodetectors based on micro-channel plates are being developed by the Large-Area Picosecond Photo Detector (LAPPD) Collaboration. These photosensors have been shown to have excellent spatial and timing resolution. Using these devices in massive water Cherenkov detectors, we could significantly improve the vertex resolution for neutrinos enhancing background rejection for neutrino oscillation experiments. We present preliminary results on the reconstruction capabilities for single particles in water Cherenkov detectors using fast photosensors.
\end{Abstract}
\vfill
\begin{Presented}
DPF 2013\\
The Meeting of the American Physical Society\\
Division of Particles and Fields\\
Santa Cruz, California, August 13--17, 2013\\
\end{Presented}
\vfill
\end{titlepage}
\def\thefootnote{\fnsymbol{footnote}}
\setcounter{footnote}{0}

\section{Introduction}

In the last few years, investigating the neutrino properties led to discoveries such as: neutrino finite mass, lepton flavor violation and oscillation with a large mixing angles. Neutrino physics is still very promising in discovery opportunities: observation of CP symmetry violation in neutrino mixing, resolution of the neutrino mass hierarchy as well as interactions with matter, searches for nucleon decay signatures, and detailed studies of neutrino bursts from galactic supernovae~\cite{NuPhysOpportunities}. Aiming for discoveries in neutrino physics requires performing measurements with high sensitivities. Neutrinos have very small interaction cross section. Thus, in order to increase the interaction probability, it is necessary to design detectors with large fiducial mass and to create a high intensity neutrino beam. Very large detectors can provide the mass required to the next generation of long-baseline, reactor and underground experiments.

\section{The next generation of neutrino experiments}

It is likely that further progress in neutrino physics can be made using large water or liquid scintillator detectors, and several proposals for such detectors are already in their planning stages: Hyper-K~\cite{hyperK},LENA~\cite{lena},LBNE~\cite{lbne}.  All the proposed detectors will attempt a broad range of physics in addition to neutrino oscillations.\\
Designing multi-purpose detectors to enable a broader physics program is a challenge, in particular when it requires very large volumes. 
The challenge of such detectors is to instrument the very large volumes and surfaces while efficiently detecting neutrinos. Developments in photosensor technology which will allow the achievement of a faster timing resolution, a better spatial granularity and a large-area coverage at lower costs, might enable new and more efficient detector designs.  
As an example, there are the two types of new large-area photosensors in development: 
\begin{itemize}
\item{Hybrid Photo-Detector - an avalanche photodiode, with the potential to achieve a time resolution between 0.6 ns and 2.2 ns, depending on the designed size~\cite{HPD}.}
\item{Large Area Pico-Second Photo-Detector (LAPPD) - developed at Argonne National Laboratory. Its design is based on micro-channel plate, with the improvement that the atomic layer deposition is used as a technique to deposit materials with different functions. These detectors have demonstrated single-photoelectron resolutions, below 100 picoseconds~\cite{LAPPD}. In addition, since micro-channel plate detectors are imaging detectors, LAPPDs also provide very good spatial resolution. This resolution is limited only by the frugal anode design of the LAPPD detectors and is currently measured to be better than a centimeter~\cite{Anode}. The ability to arrange the ${8"}\times{8"}$ tiles in a ${24"}\times{32"}$ super-module may also help in optimizing the detector coverage.}
\end{itemize}

\section{Reconstruction in water Cherenkov neutrino detectors}

\subsection{Water Cherenkov neutrino detectors}
\label{WChDetectors}
Water Cherenkov detectors often consist of a volume of water instrumented with photosensors on the surrounding surface. A shockwave of optical light (Cherenkov photons) is produced when a charged particle travels through a dielectric medium faster than the speed of light in that medium (${c/n}$), where n is the index of refraction. The photons are emitted at an angle $\theta_{C}$  relative to the track direction, and $cos{\theta_{C}=\frac{1}{n\beta}}$, where $\beta=\frac{v}{c}$. The radiation is azimuthally symmetric about the track direction, resulting in a ring-like pattern that can be identified on the photodetectors array. The quantity, spatial distribution, and arrival times of these photons provide information on the location, direction, and energy of the particle. We refer to the extraction of such information from the charge and time measurements of the photodetectors as ÔÔreconstruction.ÕÕ An analysis of the ring profile can also provide information on the identity of the particle and, if multiple rings are detected, the number of particles.

\subsection{Reconstruction using Cherenkov light}
Instrumenting large detectors approaching megaton scales with sufficient coverage to achieve the necessary physics capabilities can be prohibitively costly. Working with the faint, scattering-dominated light patterns from a low coverage detector is challenging. Some of the information loss from poor coverage may be recovered from high precision measurements of the hit positions and arrival times of those direct photons which are measured. Therefore, it may be possible to achieve better vertex resolution with less coverage, given high resolution photosensors.
%When dealing with large detectors, the light collection, the acceptance and coverage are very important in order to detect more photons, while keeping an optimum design. Distinguishing between the patterns of light can be very challenging when dealing with low photo-detectors coverage. Therefore, spatial resolution and granularity of the photo-detectors is very important. Timing resolution can also be important for better vertex reconstruction.

\subsection{Timing based vertex reconstruction algorithm}
We developed a timing-based algorithm to reconstruct the vertex in large water Cherenkov detector instrumented with LAPPDs. When the Cherenkov photons (section~\ref{WChDetectors}) propagate through the water, they attenuate and even scatter, changing the path direction, resulting in a longer traveling time. In addition, the angle $\theta_{C}$ at which the photons are emitted varies as a function of the photon wavelength. This means that depending of the wavelength, a photon can travel a larger or a shorter distance before reaching the photo-detectors on the wall. This effect of the chromatic dispersion is not negligible in large detectors. Since we are using photosensors with better timing and spatial resolution we must account for the chromatic dispersion in our reconstruction algorithm.\\
Our vertex algorithm is built around an observable distribution known as the {\bf timing-residual} distribution, which consists of the differences between the predicted and measured arrival times of the photon hits, relative to a given vertex and track hypothesis. 
%From the study of the time residual, we concluded that for a 200 kton water Cherenkov detector, a time resolution of 100 pico-seconds for the photosensors provides a sensitivity comparable to the perfect timing resolution. 
We built our PDF so that it models the shape of the time residual spectrum, taking into account the chromatic dispersion, optical attenuation and quantum efficiency. From our studies, we found that for a 200 kton detector, the photosensors with time resolution of 0.1 ns were sufficient to fully represent the shape information imbedded in the time residual distribution. The PDF was then used in a vertex reconstruction fitting algorithm based on the maximum likelihood method. 

\subsection {Results}
We simulated a 1.2 GeV muon sample traveling through a 200 kton water Cherenkov detector with 13\% photodetectors coverage. To generate the samples, we used WCSim~\cite{WCSim}, which is a Geant4 simulation validated by the LBNE collaboration. The muon scattering effect is not taken into account in the simulation.\\

\begin{figure}[!htb]
    \centering
   \includegraphics[angle=0,width=0.65\textwidth]{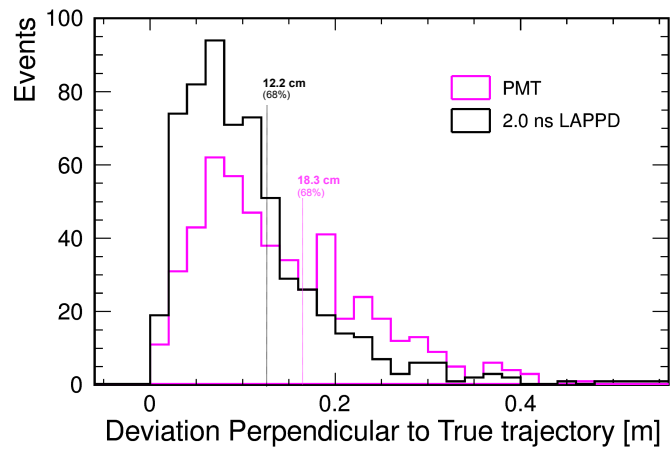}
    \caption{Difference between the position of the reconstructed vertex and the true vertex in the perpendicular direction in the case when we used conventional PMTs with 2 ns time resolution and LAPPDs with 2 ns time resolution. No muon scattering is simulated.}
   \label{fig:ungatedvsgatedPMT}
\end{figure}

Figure~\ref{fig:ungatedvsgatedPMT} shows the difference between the reconstructed vertex and the true vertex in the perpendicular direction with respect  for simulations where we used two types of photodetectors: PMT- the conventional photodetectors with 2 ns time resolution, and LAPPDs with 2 ns time resolution, respectively. The two types of photosensors have a similar time resolution, but the granularity is very different: the PMT is a single pixel detector, while the LAPPD provides the true position of the photons hitting the same photosensor. By comparing the results obtained for these cases, we concluded that using a poor time resolution and a very good granularity has only a slight improvement in reconstructing the vertex in the perpendicular direction. This agrees with the fact that the vertex reconstruction in the perpendicular direction with respect to the track direction is mainly based on the timing and minimally on the hit positions. The results were obtained for events passing a 3 m fiducial cut. The imaging of the photosensors might still be useful in reconstructing events with the vertex near the wall of the detector.   

\begin{figure}[!htb]
    \centering
   \includegraphics[angle=0,width=0.65\textwidth]{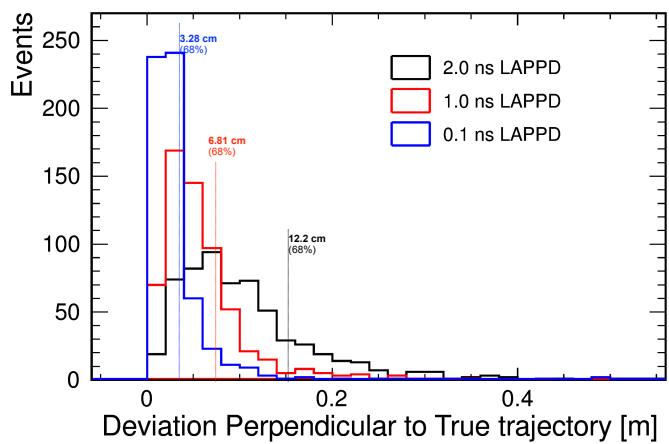}
    \caption{The distance between the reconstructed vertex and the true vertex for MCP-like with different time resolution: $0.1$ ns shown in red and 2 ns.}
   \label{fig:fitResults}
\end{figure}

Figure~\ref{fig:fitResults} shows a comparison of the distance between the reconstructed vertex and the true vertex in the perpendicular direction with respect to the track direction for the case when LAPPDs with $0.1$ ns, 1 ns and 2 ns time resolution were used. There is an observed factor of 3 improvement in the vertex reconstruction in the perpendicular direction.

\begin{figure}[!htb]
    \centering
   \includegraphics[angle=0,width=0.65\textwidth]{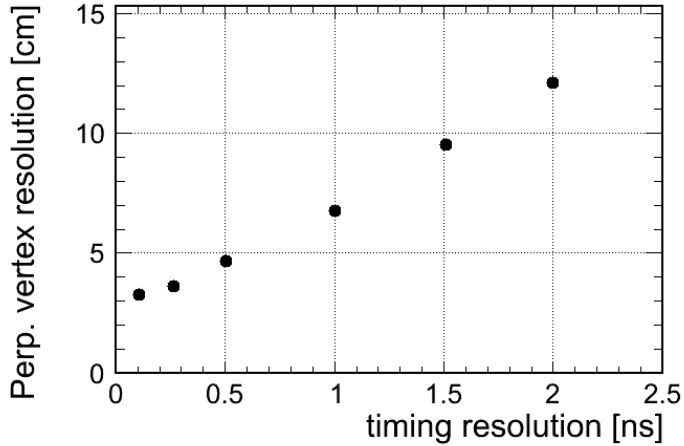}
    \caption{The position of the reconstructed vertex as a function of the time resolution of LAPPDs. This can be compared to the true vertex position, which is $= 0$.}
   \label{fig:LAPPDvtxRecoPerpDir}
\end{figure}

Figure~\ref{fig:LAPPDvtxRecoPerpDir} shows the actual position for the reconstructed vertex in the perpendicular direction as a function of the time resolution when LAPPDs with different time resolutions were used. By comparing this with the true vertex position, which is $= 0$ in this case, we can observe a trend in the vertex reconstruction, the reconstruction improving considerably when the time resolution become better and better.

\section{Conclusions}
The next generation of neutrino experiments will use very large volume, liquid-filled detectors. New photosensors such as LAPPDs could have an impact on the design of these detectors. When using photosensors with better timing and spatial resolution, we must account for chromatic dispersion, which becomes an important effect in very large detectors. We have shown that by fully exploiting the information provided by high resolution photosensors it is possible to achieve an improved vertex reconstruction. From our simulation, we concluded that by using LAPPDs with 100 psec time resolution, we achieved 3 cm vertex resolution in perpendicular direction with respect to the track direction for 1.2 GeV muons in a 200 kton detector with very low coverage.\\
The combination of fine timing and spatial resolution provides improved tracking and analysis capabilities that could be used for enhanced background rejection, energy resolution and fiducial volume definition.

\Acknowledgments
I am grateful to the entire Fast Timing group. In particular, I am grateful to Mayly Sanchez for her guidance and to Matthew Wetstein and Tian Xin for all the great work that was essential in performing this studies.

\end{document}